\documentclass[%
 aip,
 amsmath,amssymb,
 reprint,%
]{revtex4-1}

\usepackage{graphicx}% Include figure files
\usepackage{dcolumn}% Align table columns on decimal point
\usepackage{bm}% bold math
\usepackage{xcolor}
\usepackage[utf8]{inputenc}
\usepackage[T1]{fontenc}
\usepackage{mathptmx}

\begin{document}

\preprint{AIP/123-QED}

\title{Double Mpemba effect in the cooling of trapped colloids}
% Force line breaks with \\
\author{Isha Malhotra}
\email {Isha.Malhotra@hhu.de}
 
\author{Hartmut Löwen}%
\affiliation{%
Institut für Theoretische Physik II: Weiche Materie, Heinrich-Heine-Universität Düsseldorf, 40225 Düsseldorf, Germany
}%

\date{\today}% It is always \today, today,

\begin{abstract}
The Mpemba effect describes the phenomenon that a system at a hot initial temperature cools faster than at an initial warm temperature in the same environment. Such an anomalous cooling has recently been predicted and realized for trapped colloids.  Here, we investigate the freezing behavior of a passive colloidal particle by employing numerical Brownian dynamics simulations and theoretical calculations with a model that can be directly tested in experiments. During the cooling process, the colloidal particle exhibits multiple non-monotonic regimes in cooling rates, with the cooling time decreasing twice as a function of the initial temperature—an unexpected phenomenon we refer to as the Double Mpemba effect. Additionally, we demonstrate that both the Mpemba and Double Mpemba effects can be predicted by various machine learning methods, which expedite the analysis of complex, computationally intensive systems.
\end{abstract}

\maketitle

\section{\label{sec:level1}Introduction}

The Mpemba effect challenges conventional understanding by proposing that hot water can cool and freeze faster than its cooler counterpart, contrary to intuitive expectations \cite{mpemba1979mpemba}. Despite extensive experimental investigations into this phenomenon in water, a consensus regarding its underlying cause remains elusive \cite{jeng2006mpemba,wojciechowski1988freezing,vynnycky2012axisymmetric,vynnycky2015can,burridge2016questioning,auerbach1995supercooling}. Recent research advances have demonstrated that the Mpemba effect is not limited to the freezing of water but occurs in a variety of contexts. This phenomenon has been identified in granular gases \cite{lasanta2017hotter,torrente2019large,biswas2020mpemba,biswas2022mpemba,mompo2021memory,megias2022mpemba}, inertial suspensions \cite{takada2021mpemba}, Markovian models \cite{lu2017nonequilibrium,klich2019mpemba,busiello2021inducing,lin2022power}, optical resonators \cite{keller2018quenches,santos2020mpemba,patron2021strong}, spin glasses \cite{baity2019mpemba} and quantum systems\cite{chatterjee2023quantum,nava2019lindblad,carollo2021exponentially,manikandan2023autonomous,ares2023entanglement,ivander2023hyperacceleration,nava2024mpemba}. Notably, it has also been observed in colloidal particle systems undergoing rapid thermal quenching \cite{kumar2020exponentially,schwarzendahl2022anomalous}.
 In its simplest form, single particles are confined within one-dimensional asymmetric double-well potential, replicating the liquid and frozen states of water. The synthesis of experimental findings and theoretical insights, unravel the mechanisms driving this intriguing effect \cite{bechhoefer2021fresh,chetrite2021metastable,lu2017nonequilibrium}, thereby advancing our comprehension of its fundamental principles.

\begin{figure}[h!]
 \centering
	\includegraphics[scale=0.16]{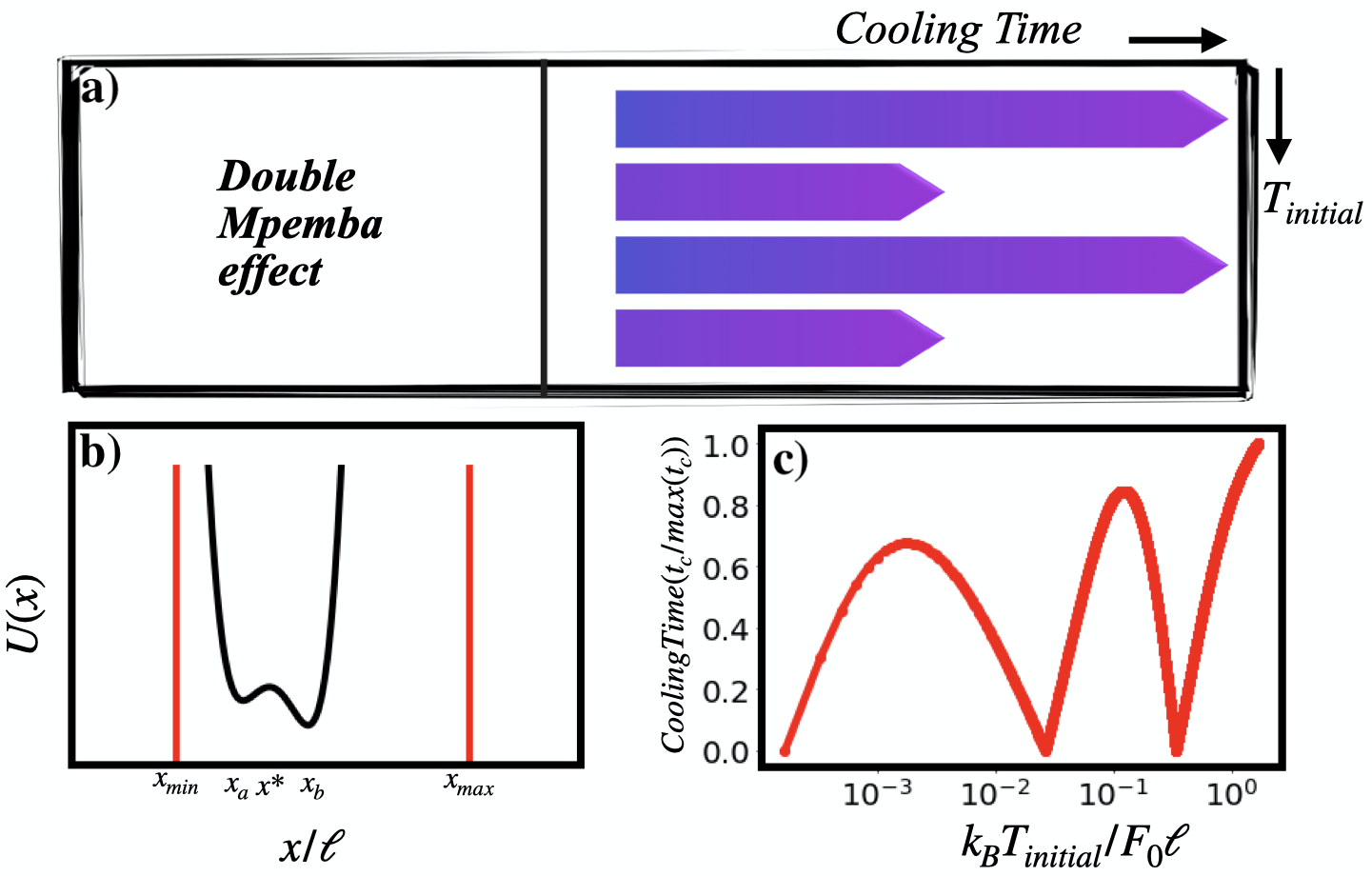}
	\caption{a) Double Mpemba effect, where each colored arrow represents a cooling process and the arrow's length depicts the time the system needs to cool down. b) Asymmetric potential $U(x)$ with repulsive walls at $x_{min}$ and $x_{max}$, potential minimal at $x_a$ and $x_b$ and a maxima at $x^*$. c) Cooling time $t_c$ as a function of initial temperature to a fixed bath temperature $T_{bath}$.}
 \label{Fig1}
 \end{figure}

In this study, we examine the cooling process of a trapped colloid within a potential featuring two repulsive walls shown in Fig.~\ref{Fig1}b and discover that it exhibits a pronounced Mpemba effect, occurring not just once but twice if the initial temperature is varied (Fig.~\ref{Fig1}a and~\ref{Fig1}c) a phenomenon which we report as \textit{Double Mpemba effect}. Furthermore, we explore how imposed bath temperatures influence the type of Mpemba -normal, or Double - that the system exhibits. We have generalized a simple theoretical framework proposed by Kumar \textit{et al.}\cite{kumar2020exponentially} that explains the observations of numerical simulations and quantitatively agrees with the analysis based on the eigenfunction expansion of the Fokker-Planck equation \cite{schwarzendahl2022anomalous,kumar2020exponentially,lu2017nonequilibrium,sandev2015diffusion}. Furthermore, traditional experimental and computational approaches to studying the Mpemba effect often face challenges due to the complexity and variability of the parameters involved. To overcome, these challenges, we propose a novel approach that leverages theoretical modeling and machine learning \cite{amorim2023predicting,cunningham2008supervised,hastie2009overview,breiman1984classification} to predict the colloidal Mpemba effect with high accuracy.

To illustrate the Mpemba effect, imagine two systems with temperatures ranging from warm to hot. Typically, when these systems are cooled to a set cold bath temperature, we would expect that the hotter the system, the longer it would take to cool. However, the Mpemba effect occurs when the hot system cools faster than the warm one. In the case of a passive colloid in an asymmetrical potential, this happens because the hot particle has enough residual energy to overcome the barrier and quickly settles into the cold state. In contrast, a warm particle, with less residual energy, takes longer to cross the barrier. We show the existence of the Double Mpemba effect and that the key factors influencing the Mpemba effect are not just the residual energy but also the initial state of the system and the final bath temperature. This finding broadens our understanding of the Mpemba effect and highlights the complexity of cooling dynamics in these systems.

\section{\label{sec:model}Model and simulation technique}
We explore the process of cooling for a Brownian colloidal particle confined within a double well potential through numerical simulations. The symmetry of the double well potential is broken either by bringing a tilt in the potential or by the asymmetric placement of the potential in a domain (see Fig. 1b).  The motion of the Brownian particle, experiencing fluctuations at temperature $T$ and undergoing overdamped motion, is described in one spatial dimension by the equation:

\begin{equation}
\frac{dx}{dt} = -\frac{1}{\gamma}\partial_x U(x) + \eta(t)
\label{eqn1}
\end{equation}
where $\eta(t)$ represents Gaussian white noise with zero mean and variance $\left<\eta(t)\eta(t^{'})\right> = 2 D_T \delta(t-t^{'})$. Here, the noise strength corresponds to the translational diffusion constant $D_T$ of the particle, which is determined by the temperature $T$, given by the Stokes-Einstein relation $D_T = \frac{k_B T}{\gamma}$, where $k_B$ denotes the Boltzmann constant and $\gamma$ represents the friction coefficient. The particle is subjected to an external double well potential similar as in \cite{schwarzendahl2022anomalous,kumar2022anomalous} defined as follows:

\begin{equation}
U(x) = \begin{cases}
-F_0 x &\text{if } x<x_{min}\\
F_1 [(1-x^2)^2 - 0.5 x] &\text{if } x_{min}<x<x_{max}\\
F_0 x &\text{if } x>x_{max}
\end{cases}
\label{eqn2}
\end{equation}

The components in Eq.~\ref{eqn2} that scale with $F_0$ signify the presence of repulsive barriers positioned at $x_{min}$ and $x_{max}$ such that the forces are constant beyond $x_{min}$ and $x_{max}$, while the component proportional to $F_1$ describes an asymmetric potential featuring two minima at $x_a$ and $x_b$ of varying heights and maxima at $x^*$. 

The length of the confining box denoted as \( \ell = |x_{max} - x_{min}| \), serves as a convenient unit of length. When this is combined with the translational diffusion constant, it yields a natural time scale expressed as \(\tau_D = \frac{\ell^2}{D_T}\). Throughout this paper, the temperatures are defined in units of $F_0\ell$.

To gain quantitative insight into the relaxation process, we quantify the distance between the target equilibrium distribution $\pi_{\text{bath}}(x)$ and the probability distribution $P(x, t)$ of a particle generated from Eq.~\ref{eqn1} during the cooling process \cite{schwarzendahl2022anomalous,lu2017nonequilibrium}. To construct this distance measure, we discretize the spatial components of both $\pi_{\text{bath}}(x)$ and $P(x, t)$ into $N$ grid points, resulting in $\pi_{i, \text{bath}}$ and $P_i(t)$, respectively. The distance measure is then defined as:
\begin{equation}
\mathcal{D}(t) = \frac{1}{N} \sum_{i=0}^{N} |P_i(t) - \pi_{i, \text{bath}}|.
\label{DT}
\end{equation}

 In the following, we present a theoretical formula by generalizing the approach proposed by Kumar \textit{et al.} for calculating the cooling time scale of particles starting from various initial temperatures $T_{initial}$. The occupation ratios/probabilities $N_a(T)$ and $N_b(T)$, which indicate the probability of a particle in the left-hand domain $(-\infty, x^*)$ and right-hand domain $(x^*, \infty)$ respectively, at a temperature $T$ with $\beta=\frac{1}{k_B T}$ in equilibrium is given as:
\begin{eqnarray}
 N_{a}(T) = \frac{\int_{-\infty}^{x^*} \exp(-\beta U_a(x)) \,dx }{\int_{-\infty}^{\infty} \exp(-\beta U_a(x)) \,dx }
 \end{eqnarray}
 \begin{eqnarray}
  N_{b}(T) = 1- N_{a}(T) 
\end{eqnarray}

\noindent The time scale for cooling is approximately given as \cite{russel1991colloidal,Chandrasekhar1943}:
 \begin{eqnarray}
     t_c  \approx \tau_D |N_a(T_{initial}) - N_a(T_{bath})| \nonumber\\
    \times \, \exp\left(\frac{\Delta E_i}{k_B T_{bath}}\right)
    \label{timescale}
 \end{eqnarray}

 \noindent The Arrhenius-like exponential factor accounts for the diffusion over an energy barrier that a particle originally in the potential hole at $x_a$, will escape to $x_b$ crossing the barrier at $x^{*}$ (see Fig.~\ref{Fig1}b)\cite{hanggi1990reaction,Chandrasekhar1943,kraikivski2004barrier,sharma2017escape,scacchi2019escape}. The expressions $\Delta E_a = U(x^*) - U(x_a)$ and $\Delta E_b = U(x^*) - U(x_b)$ define the energy barriers for the potential minima at $x_a$ and $x_b$, respectively. Here, $\Delta E_i$ is equal to $\Delta E_a$ if $N_a(T_{initial}) > N_a(T_{bath})$, and it is equal to $\Delta E_b$ otherwise. 
 
 \section{\label{sec:result}Results}
 
\begin{figure}[h!]
 \centering
	\includegraphics[scale=0.14]{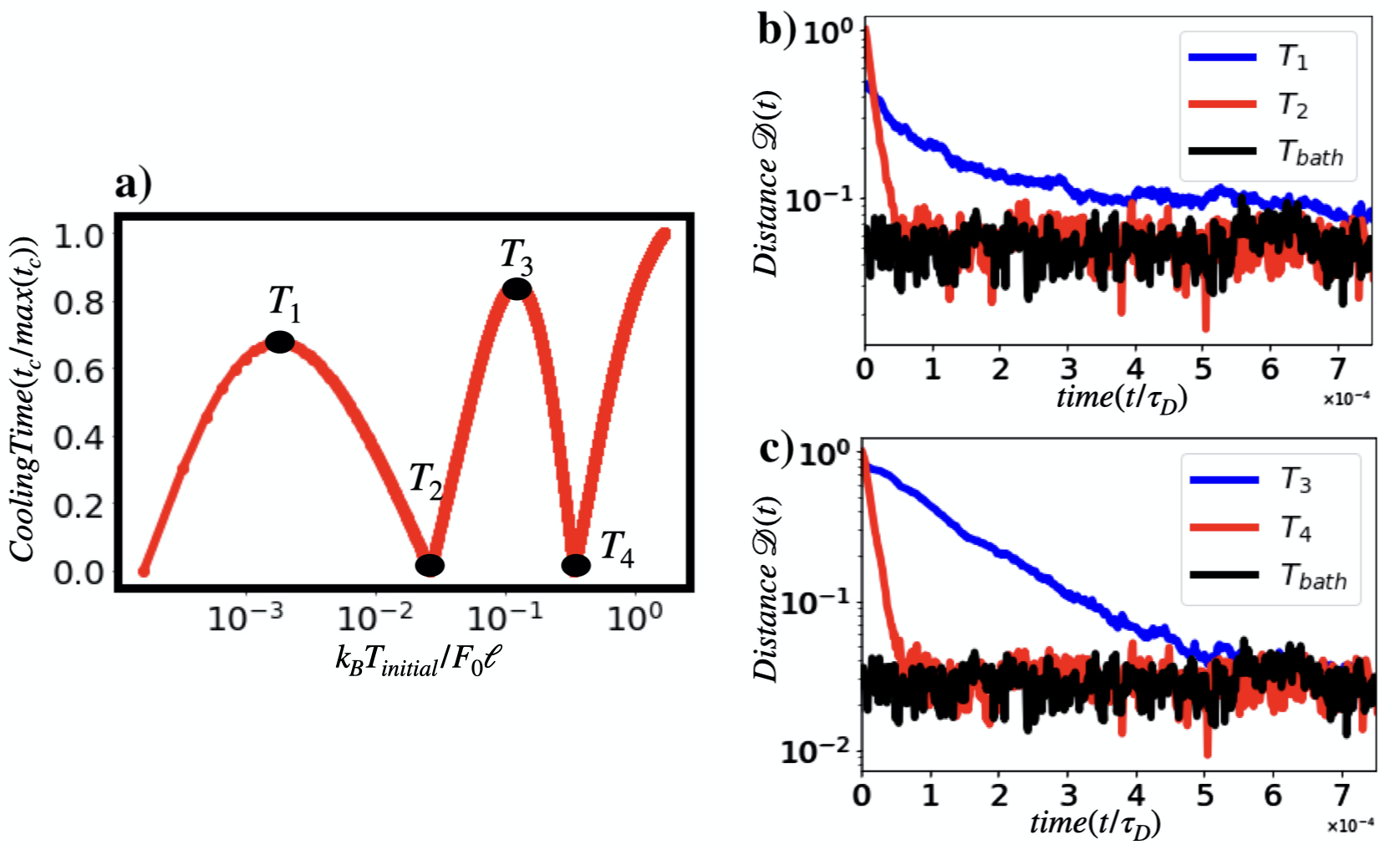}
	\caption{a) Theoretical cooling time $t_c$ as a function of initial temperature for a fixed final temperature $T_{bath}$. b) Relaxation dynamics of the 'distance' $\mathcal D(t)$, measured by comparing the probability distributions with the steady-state distribution of a cold colloid, for initially hot, warm, and cold colloid states.}
 \label{Fig2}
 \end{figure}

 In Fig.~\ref{Fig2}a, we present the cooling curve calculated from the theory as a function of different initial temperatures. The cooling time $t_c$ has a double minima, indicating the presence of the Double Mpemba effect for our chosen parameters. Numerical simulations further confirm theoretical predictions where we calculate the distance measure $\mathcal D(t)$ as defined in Eq.~\ref{DT} in Figs. 2b and 2c. From this measure, we can extract a cooling time $t_c^{sim}$, defined as the time at which $\mathcal D(t)$ has decayed to zero or, in our case, to the noise level. We show that particles at temperatures $T_2$ and $T_4$ cool very quickly, while particles at temperatures $T_1$ and $T_3$ take longer to relax, fully consistent with the theoretical calculations.

\begin{figure}[h!]
 \centering
	\includegraphics[scale=0.18]{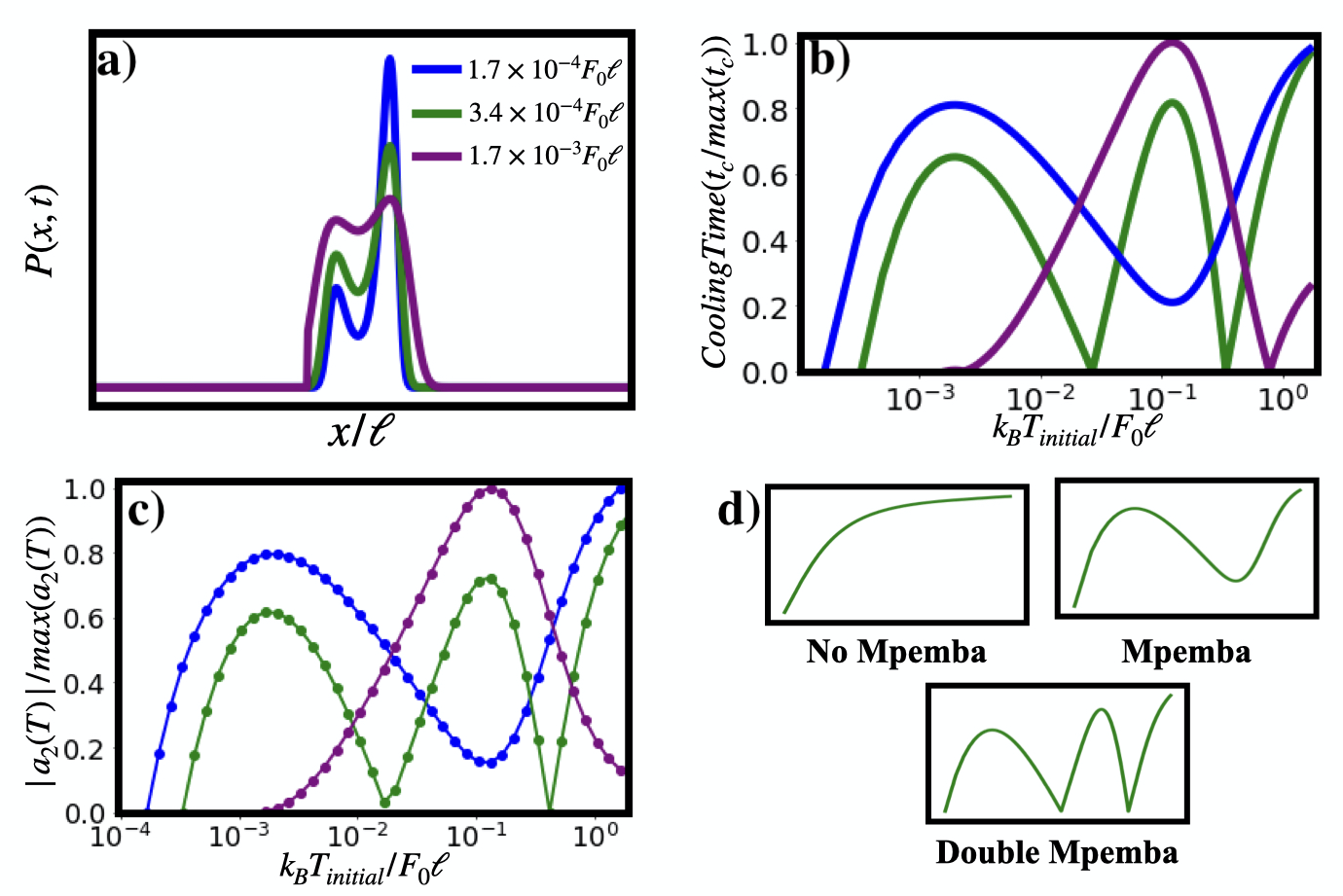}
	\caption{a) Probability distributions of colloidal particles at different $T_{bath}$ temperatures as indicated in the figure. b) Cooling time normalized to [0,1] by the maximum value of cooling time ($\max(t_c)$) in the temperature range $10^{-4} F_0 \ell$ to $3 \times 10^0 F_0 \ell$ as a function of initial temperature for different bath temperatures calculated from the theory. c) The normalized second eigenvalue coefficient \( \frac{|a_2(T)|}{\max(a_2(T))} \) as a function of the initial temperature \( T_{initial} \), is obtained by numerical calculations based on the Fokker Planck equation. d) Variation of cooling time as a function of initial temperature illustrating No Mpemba, Mpemba, and Double Mpemba effect. }
 \label{fig3}
 \end{figure}

 \begin{figure}[h!]
 \centering
	\includegraphics[scale=0.12]{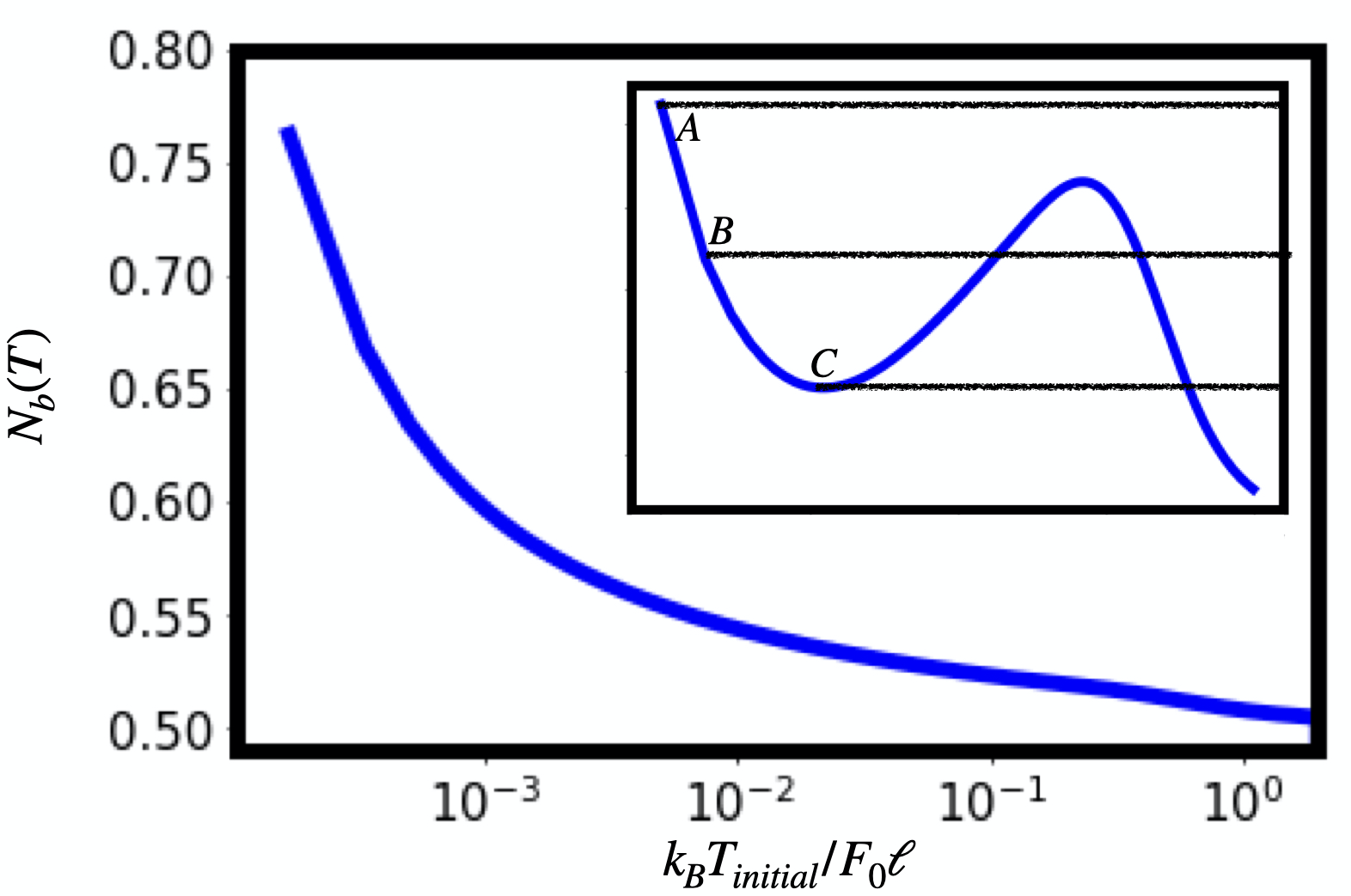}
	\caption{a) The occupation probability $N_b(T)$ as a function of temperature $T$ for a system exhibiting No Mpemba effect. [Parameters: $x_{min}= -0.5\ell, x_{max}=0.5\ell, F_0=750, F_1 = 0.0016 F_0$]. Inset shows $N_b(T)$ as a function of temperature $T$ for a system exhibiting Mpemba $(A$ and $C)$ and Double Mpemba effect $(B)$. [Parameters: $x_{min}= -0.25\ell, x_{max}=0.75\ell, F_0=750, F_1 =0.0016 F_0$].  }
 \label{Nb}
 \end{figure}

To understand the effect of different bath temperatures, we present the particle distribution and calculate the normalized cooling time from the theoretical model at various bath temperatures, as shown in Fig. \ref{fig3}a and Fig. \ref{fig3}b, respectively. We observe that at \( T_{\text{bath}} = 1.7 \times 10
^{-4} F_0 \ell\), the system exhibits the normal Mpemba effect, whereas at \( T_{\text{bath}} = 1.7 \times 10
^{-3} F_0 \ell \), a strong Mpemba effect is observed. At \( T_{\text{bath}} = 3.4 \times 10
^{-4} F_0 \ell \), the cooling time shows double minima, indicating the presence of the Double Mpemba effect.

To validate the theoretical model and numerical simulations, we employ a recent approach that relates the Mpemba effect to an eigenvalue expansion \cite{bechhoefer2021fresh, schwarzendahl2022anomalous, lu2017nonequilibrium}. 
The probability density $p(x,t)$ can be represented as an infinite sum of eigenfunctions of the Fokker-Planck equation, which governs its evolution. The theory subsequently predicts that the density function is primarily influenced by the first two terms of the infinite series at long times.
\begin{equation}
p(x,t) \approx \pi(x;T_{bath}) + a_2(T) v_2(x) e^{-\lambda_2 t}
\end{equation}
where the coefficients $a_2(T)$ is a real number that depends on the initial temperature and the potential energy. This approach shows that the cooling time at different initial temperatures is proportional to the second eigenvalue coefficient \( |a_2(T)| \) of the Fokker-Planck equation. In Fig. \ref{fig3}c, we display the normalized \( |a_2(T)| \) at different bath temperatures and demonstrate that the cooling times in Fig. \ref{fig3}b quantitatively agree with the values of \( |a_2(T)| \). Finally, in Fig. \ref{fig3}d, we illustrate the cooling time plots for different types of Mpemba effects.

 To understand the origin of different Mpemba effects, we examine the role of the first term defined in Eq.~\ref{timescale}. In cases where the initial occupation probability matches the occupation probability at the final bath temperature, the particle relaxes very quickly as it does not need to cross the barrier to hop from the potential well at $x_a$ to $x_b$. In Fig.~\ref{Nb}, we show the occupation probability $N_b(T)$, which indicates the likelihood of a particle being in the right-hand domain $(x^{*}, \infty)$ as a function of initial temperature $T_{initial}$. If for a given $T_{bath}$, $N_b(T) = N_b(T_{bath})$, the particle immediately relaxes to the cold distribution, resulting in a pronounced Mpemba effect. Conversely, if $N_b(T) \neq N_b(T_{bath})$, the particle must overcome the potential barrier to reach the equilibrium distribution at $T_{bath}$. Since crossing this barrier takes a considerable amount of time, the particle relaxes slowly. In Fig. \ref{Nb}, we observe that $N_b(T)$ decreases monotonically with increasing temperature for the given parameters as indicated in the caption. This implies that hotter particles will take longer to relax to the cold distribution compared to warmer particles, indicating the absence of the Mpemba effect. Conversely, the inset displays $N_b(T)$ with non-monotonic behavior for a set of parameters mentioned in the caption, suggesting the presence of Mpemba, Double Mpemba, and strong Mpemba effects at $A$, $B$, and $C$, respectively. The Mpemba effect is characterized by the non-monotonic behavior of $N_b(T)$, and a strong Mpemba effect is observed when $N_b(T) = N_b(T_{bath})$. For the Double Mpemba effect, there are two temperatures where $N_b(T) = N_b(T_{bath})$, indicating a rapid transition to the cold distribution twice during the cooling process.
\begin{figure}[h!]
 \centering
	\includegraphics[scale=0.14]{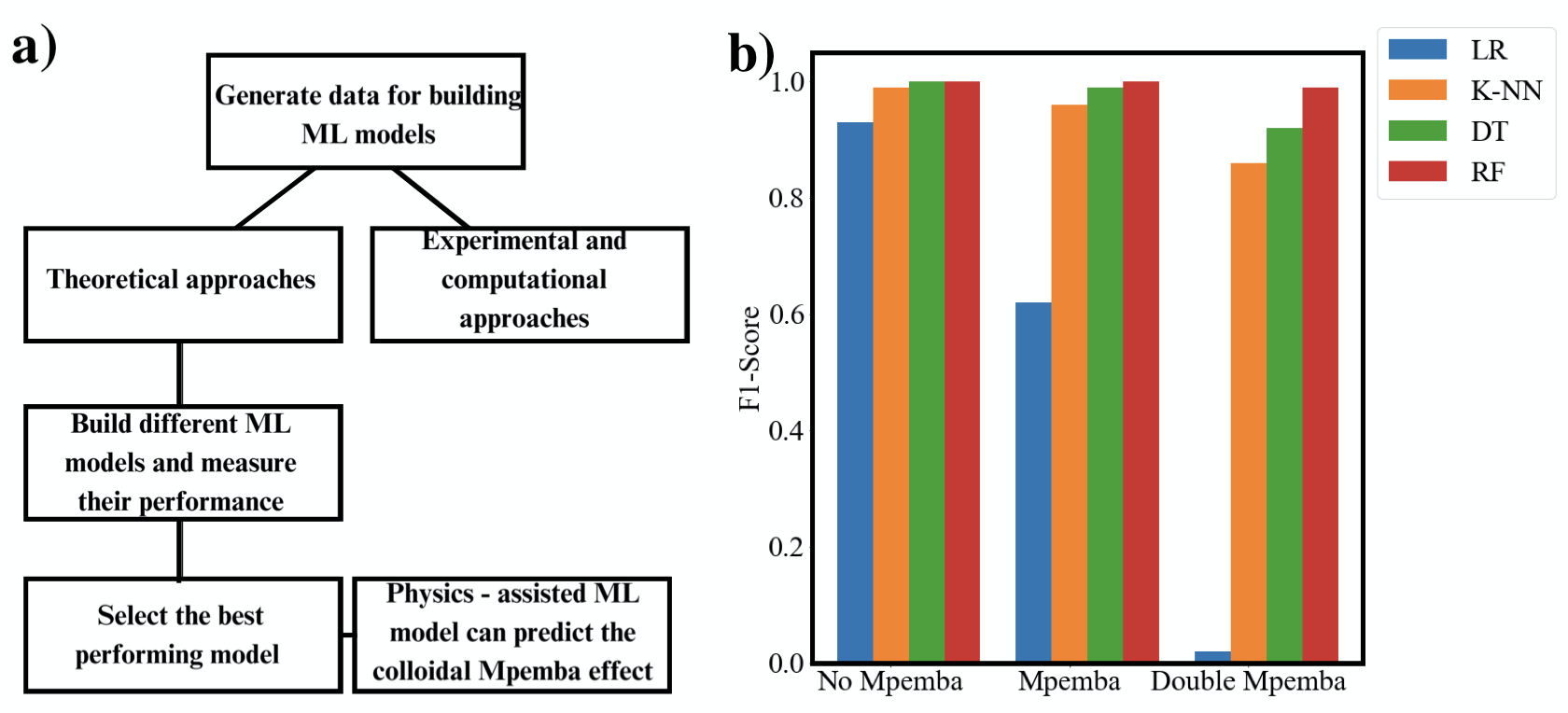}
	\caption{a) Overview of the methodology for predicting the Mpemba effect. The process begins with generating data for building machine learning models either by theoretical or experimental/computational approaches. This data is used to build and evaluate multiple ML models. The performance of these models is assessed, and the best-performing model is selected. Finally, the selected physics-assisted machine learning model is used to predict the colloidal Mpemba effect. b) Calculated F1-score of different ML models in predicting No-Mpemba, Mpemba, and Double Mpemba.}
 \label{Fig5}
 \end{figure}

Finally, we employ Machine learning (ML) to study the Mpemba effect without relying on computationally intensive calculations of eigenvectors or cooling times across varying initial temperatures. Utilizing machine learning algorithms, we can effectively classify and predict different Mpemba effect types solely based on observed data patterns. Initially, we establish a robust training dataset using the previously described theoretical model, which includes parameters such as potential and bath temperature. This dataset is structured for subsequent analysis as follows.

\begin{equation}
\left\{ x_{min}, x_{max}, T_{bath}, F_0, F_1 \right\} -\rangle
\begin{cases}
0 & \text{if No Mpemba} \\
1 & \text{if Mpemba} \\
2 & \text{if Double Mpemba} \\

\end{cases}
\end{equation}
By systematically varying these parameters, we generate a robust dataset that encompasses a wide range of scenarios. This dataset is then utilized to train multiple machine learning models, including logistic regression (LR) \cite{Cramer2004}, decision trees (DT) \cite{quinlan1986induction}, random forests (RF) \cite{Breiman2001}, and K-nearest neighbors (K-NN) \cite{Devijver1982}. The performance of these models is evaluated using the F-1 score\cite{sokolova2006beyond}, a metric that balances precision and recall to provide a comprehensive measure of model accuracy. The detailed process is illustrated in Fig \ref{Fig5}a. Initially, the dataset is generated through theoretical models, computational simulations, or experiments. In the current work, we have used a theoretical model detailed previously. Subsequently, various machine learning models are built and their performances are measured. The best-performing model is then selected to predict the Mpemba effect under different conditions. In Fig \ref{Fig5}b, we present the F1-scores of different machine learning models in predicting the No Mpemba, Mpemba, and Double Mpemba effects. The random forest exhibits the highest F-1 score among these models across all three Mpemba scenarios.

\section{\label{sec:conclusion}Conclusions}
We have investigated the influence of potential parameters and bath temperature on the manifestation of different types of Mpemba effects, demonstrating how these factors can fundamentally alter the relaxation process and lead to the Double Mpemba effect, characterized by a cooling trajectory with two minima. Furthermore, we generalized a simple theoretical framework that quantitatively aligns with the analysis based on the eigenfunction expansion of the Fokker-Planck equation \cite{kumar2020exponentially}. Additionally, we have integrated our theoretical model with advanced machine-learning techniques to enhance the predictability of this intriguing phenomenon. 

Future research could explore the application of our findings to other systems exhibiting the Mpemba effect. It would be particularly interesting to examine how varying bath temperatures and different types of potentials affect the Mpemba effect in systems such as active colloids and many-particle systems. This model can also be used to study the Mpemba effect in quantum systems \cite{chatterjee2023quantum, nava2019lindblad}, offering insights into the relaxation dynamics and thermal behaviors of complex quantum systems. A promising avenue for future research is to deepen our investigation into the Kovacs effect using our model\cite{kovacs1963dynamic,kovacs1979isobaric,bertin2003kovacs,kursten2017giant,santos2024mpemba}, aiming to uncover its intricate dynamics and implications in diverse physical systems.

Our results can be tested in real-space experiments of colloidal particles in an optical double-well potential \cite{kumar2020exponentially,buttinoni2022active}. To study the effect of different initial temperatures $T_{initial}$, the external potential has to be switched from one initial to another, which is proportional to the initial one. 

\section{Acknowledgements}
I.M. acknowledges support from the Alexander von Humboldt Foundation. H.L. acknowledges support from the Deutsche Forschungsgemeinschaft (DFG) within the project LO 418/29.

\bibliographystyle{jcp} % <===========================================
\bibliography{sorsamp} % to use file created by filecontents ...

\end{document}